\documentclass[a4paper,12pt]{article}
\usepackage[dvips]{graphicx}
\usepackage{amsmath}
\usepackage{latexsym}
\begin{document}

\title{
Behavior of Hadrons at Finite Density  \\
{\Large -- Lattice Study of Color SU(2) QCD -- }
}

\author{
Shin Muroya~$^{1)}$, Atsushi Nakamura~$^{2)}$ and Chiho Nonaka~$^{3)}$
\vspace{.3cm} 
\\
{\it $^{1)}$~Tokuyama Women's Coll. Tokuyama, 745-8511, Japan}
\\
{\it $^{2)}$~RIISE, Hiroshima Univ.,  Higashi-Hiroshima 739-8521, Japan}
\\
{\it $^{3)}$~Dept.\ of Phys., Duke Univ.,  Durham, NC27708-0305, USA}
}

\maketitle

\abstract{
Using two-color lattice QCD with Wilson fermions,  
we report a study of the finite baryon number density system
with two-flavors. 
First we investigate the Polyakov line and thermodynamical quantities
in the $(\kappa,\mu)$ plane,
where $\kappa$ and $\mu$ are the hopping parameter and chemical
potential in the fermion action, respectively.
Then we calculate propagators of meson ($\bar{q}\Gamma q$) and 
baryon ($q\Gamma q$) states.
We find that the vector meson propagators are strongly modified
in large $\mu$ regions, indicating the reduction of the mass.
This anomalous behavior of the vector meson is observed for the first 
time in lattice QCD.
\bigskip
\\
\noindent
{\it PACS:} 12.38Gc, 11.15Ha, 12.38Mh, 24.85+p
\\
\noindent
{\it Keywors: Lattice gauge theory, Finite density QCD} 
}

\section{Introduction}

QCD has confinement and deconfinement phases when the temperature,
$T$, is varied.
When an additional parameter, the density or the chemical 
potential, is added, QCD may have a much richer structure \cite{CCC,Kunihiro}.
Experimentally, a wide region of the $(T,\mu)$ plane has been 
investigated by AGS, SPS and RHIC, 
and a higher density realm might be covered by GSI and
JHF in the future. There have been experiments which suggest that vector
meson masses are modified at finite density \cite{CERES, Enyo}.

Lattice QCD
is expected to provide nonperturbative information of the hadronic world at
finite density as the first principle calculation \cite{Karsch-Hasenfratz}.
However, numerical study of lattice QCD with chemical potential is
extremely difficult, because at finite $\mu$, the fermion matrix
does not satisfy the usual condition, 
\begin{equation}
   D^\dagger = \gamma_5 D \gamma_5,
\label{D-real}
\end{equation}
and the fermion determinant $\mbox{det}D$ becomes complex,
which appears in the Euclidean path integral measure, 
\begin{equation}
Z = \int \mathcal{D}U \mathcal{D}\bar{\psi}\mathcal{D}\psi 
    \, e^{-\beta S_G - \bar{\psi}D\psi}
 = \int \mathcal{D}U \mbox{det}D \, e^{-\beta S_G} .
\end{equation}

In order to circumvent the above difficulty, 
two-color QCD has been investigated 
\cite{Nakamura84, Hands99,Hands00,KTSin,Kogut01,Kogut02,Lomba02}.  
Although recent progress of lattice calculations with the finite chemical
potential has been prominent \cite{Taro01, Fodor, Ejiri, Philipsen, DElia},
it is still very difficult to study regions around critical $\mu$
at low temperature by lattice QCD simulations. 
We report here, for the first time, 
a lattice study of hadron propagators together 
with thermodynamical quantities for
finite density SU(2) QCD with Wilson fermions.
Two-flavor case is studied.

The two-color SU(2) theory qualitatively has most of the important features 
of QCD, 
such as deconfinement transition at finite temperature.
The essential difference between two- and three-color systems is that 
in the SU(2) case baryons are made of two quarks, i.e., they are bosons.  
Therefore we must be careful in the study of two-color QCD 
in order to gain some insight
into a subset of the phenomena expected in QCD. It is indispensable
to compare results for mesons and diquarks which have a special relation
in the SU(2) case. 
Recent theoretical progress has greatly improved our understanding of
the two-color QCD system 
\cite{Schaefer,KSteT,KSTVZ,NBI}.

\section{Actions and algorithm}

We introduce the chemical potential in the conventional manner,
\begin{eqnarray}
D(x,x') = \delta_{x,x'}
 - \kappa \sum_{i=1}^{3} \left\{ 
        (1-\gamma_i) U_i(x) \delta_{x',x+\hat{i}}
      + (1+\gamma_i) U_i^{\dagger}(x') \delta_{x',x-\hat{i}} \right\}
\nonumber 
\\
  - \kappa \left\{ 
        e^{+\mu}(1-\gamma_4) U_4(x) \delta_{x',x+\hat{4}} 
      + e^{-\mu}(1+\gamma_4) U_4^{\dagger}(x') \delta_{x',x-\hat{4}} 
\right\} .
\label{Wfermion}
\end{eqnarray} \noindent
For the gauge action, we employ the plaquette and Iwasaki improved actions.
Here, we report the improved action case only, and
for analyses with the plaquette gauge action, we refer to Ref.~\cite{MNN00}.
Little is known about dynamical fermion simulations
in which the chemical potential is introduced.  We therefore
employ an algorithm where the ratio of the determinant,
\begin{equation}
 \frac{ \mbox{det} D(U+\Delta U) }{ \mbox{det} D(U) } 
= \mbox{det} (I + D(U)^{-1} \Delta D)
\end{equation} \noindent
is evaluated explicitly in each Metropolis update process, 
$U \rightarrow U+\Delta U$, 
where 
$\Delta D \equiv D(U+\Delta U) - D(U)$.  
For details of the algorithm, see Refs.\cite{Barbour87,Nakamura88}.
This algorithm has a long Markov step and is very reliable \cite{Taro87}.
Numerical costs are, however, huge and we are restricted to
small lattices. 
In the following studies, therefore, we check that the results obtained
are not sensitive to the boundary conditions.

\section{Study of the $(\kappa,\mu)$ Parameter Space}

Since there are few color $SU(2)$ lattice studies using Wilson fermions 
with finite $\mu$ in the literature except Ref.\cite{Nakamura84},
we first investigate the relevant parameter space. 
We measure the Polyakov line, $\langle L \rangle$ on a $4^4$ lattice by changing
$\beta$ for $\mu=0$ and $\kappa=0.150$ and choose the region where
$\langle L \rangle$ is small, 
i.e., the system is in the confinement phase at zero baryon number density.  
We set $\beta=0.7$ on the basis of this analysis.  

\begin{center}
\begin{figure}[thb]

\begin{center}
\begin{minipage}{ 0.4\linewidth}
\includegraphics[width= 0.9\linewidth]{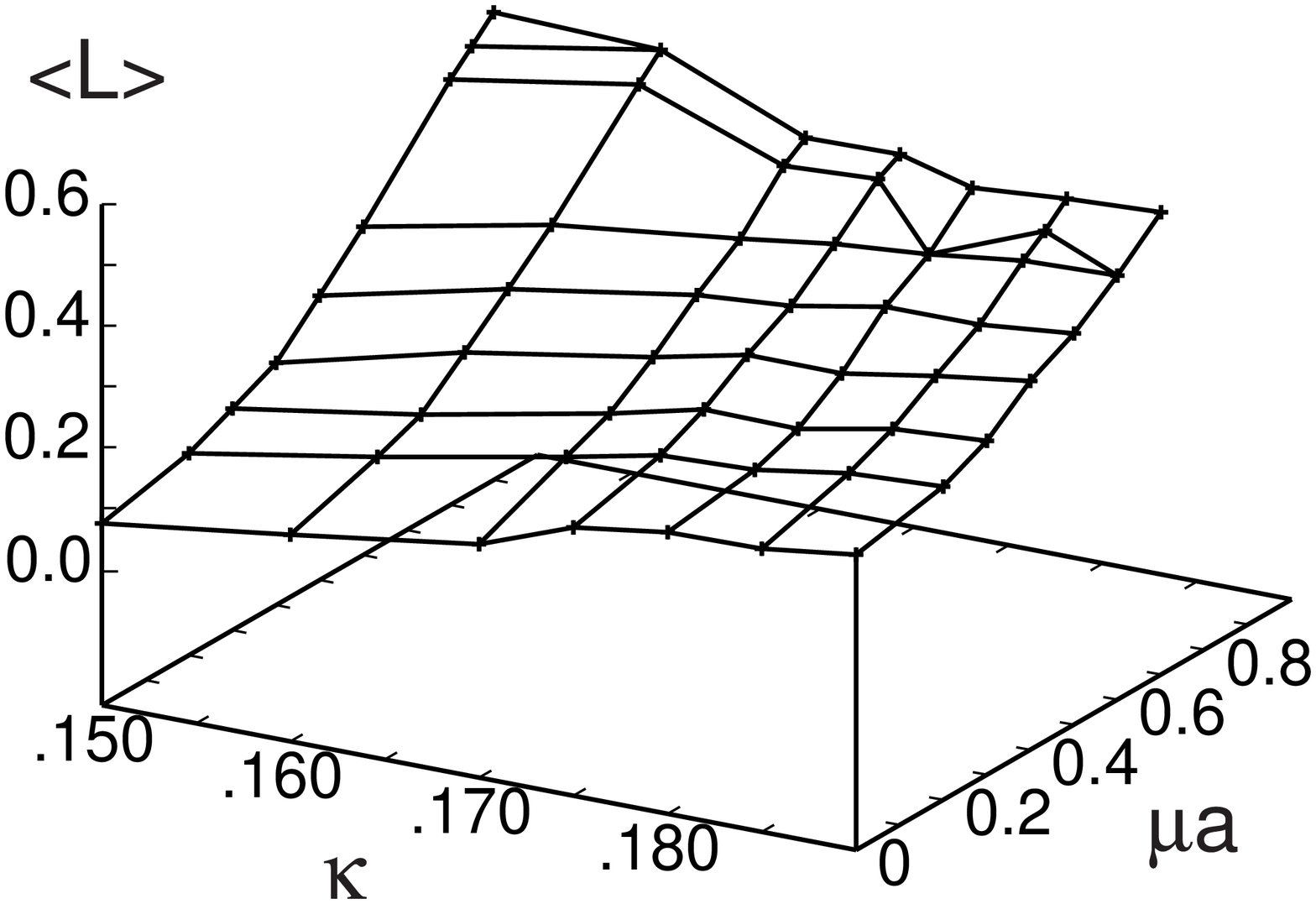}
\end{minipage}
\begin{minipage}{ 0.4\linewidth}
\includegraphics[width= 0.9\linewidth]{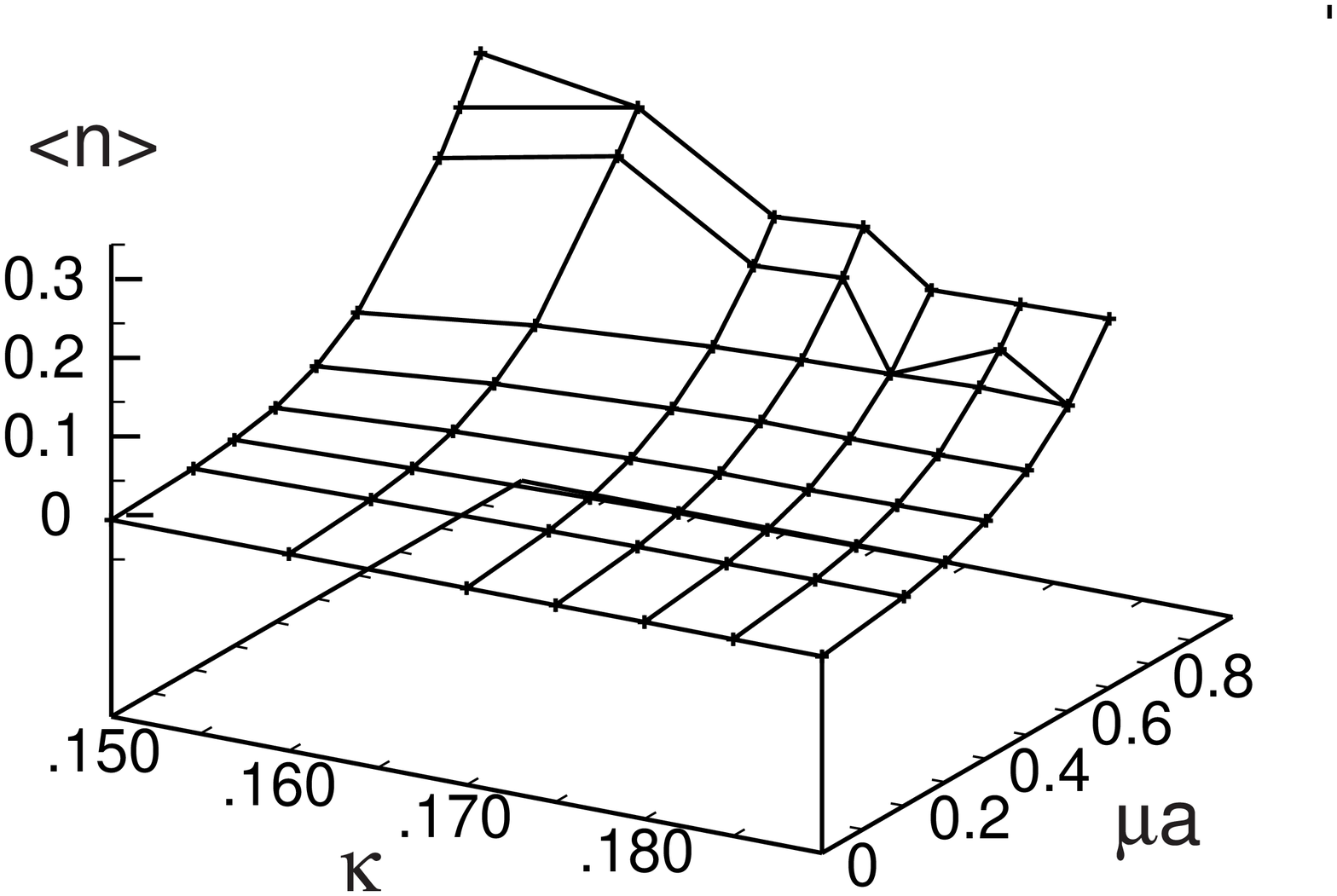}
\end{minipage}
\end{center}

\begin{center}
\begin{minipage}{ 0.4\linewidth}
\includegraphics[width= 0.9\linewidth]{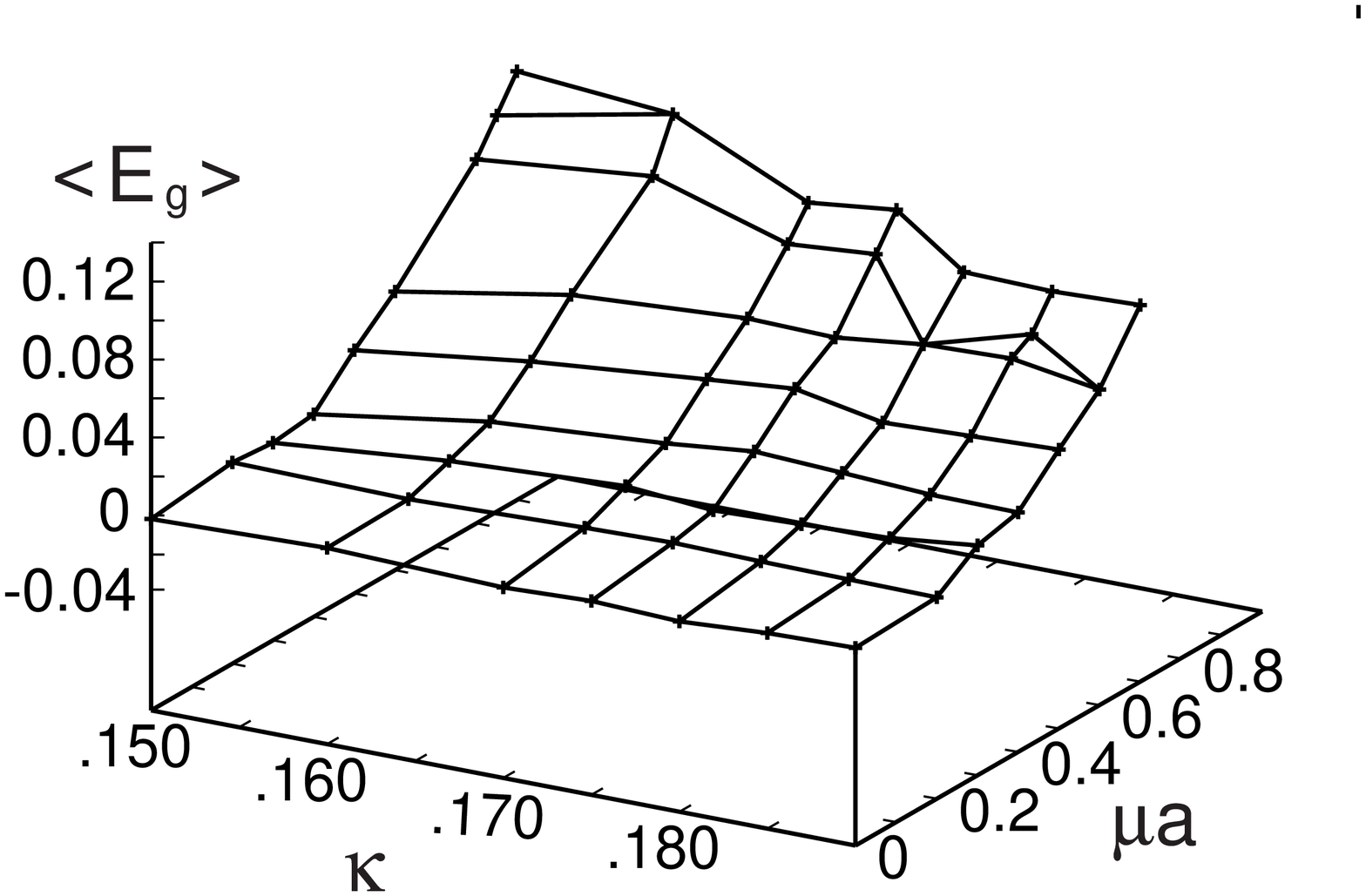}
\end{minipage}
\end{center}

\caption{
The Polyakov line $\langle L \rangle$, the number density $\langle n \rangle$ and 
the gluon energy $\langle E_g\rangle$ as a function of $\kappa$ and
$\mu$. Lattice size is $4^4$.
}
\label{Fig3D}
\end{figure}
\end{center}

At this value of $\beta$, we measure $\langle L \rangle$, 
its susceptibility, $\frac{\partial \langle L \rangle}{\partial \mu}$,
the gluon energy density, 
$\langle E_g\rangle =<\frac{1}{V_s}\frac{\partial}{\partial (1/T)}S_G>$ 
and the number density, 
$\langle n \rangle =\frac{T}{V_s}\frac{\partial}{\partial\mu}\mbox{log}Z$,
as a function of $\mu$ and $\kappa$.
Here $S_G$ is the gauge action and
$V_s=N_x N_y N_z$ is the spatial volume of the lattice.
In Fig.\ref{Fig3D}, we show $\langle L \rangle$, $\langle n \rangle$ and 
$\langle E_g \rangle$ 
on a $4^4$ lattice as a function of $\mu$ and $\kappa$.
They increase as $\mu$ becomes large and show the deconfinement
behavior.
We observe that the simulation always breaks down when we increase $\mu$ 
further.

\begin{figure}[thb]

\begin{center}
\begin{minipage}{ 0.4\linewidth}
\includegraphics[width= \linewidth]{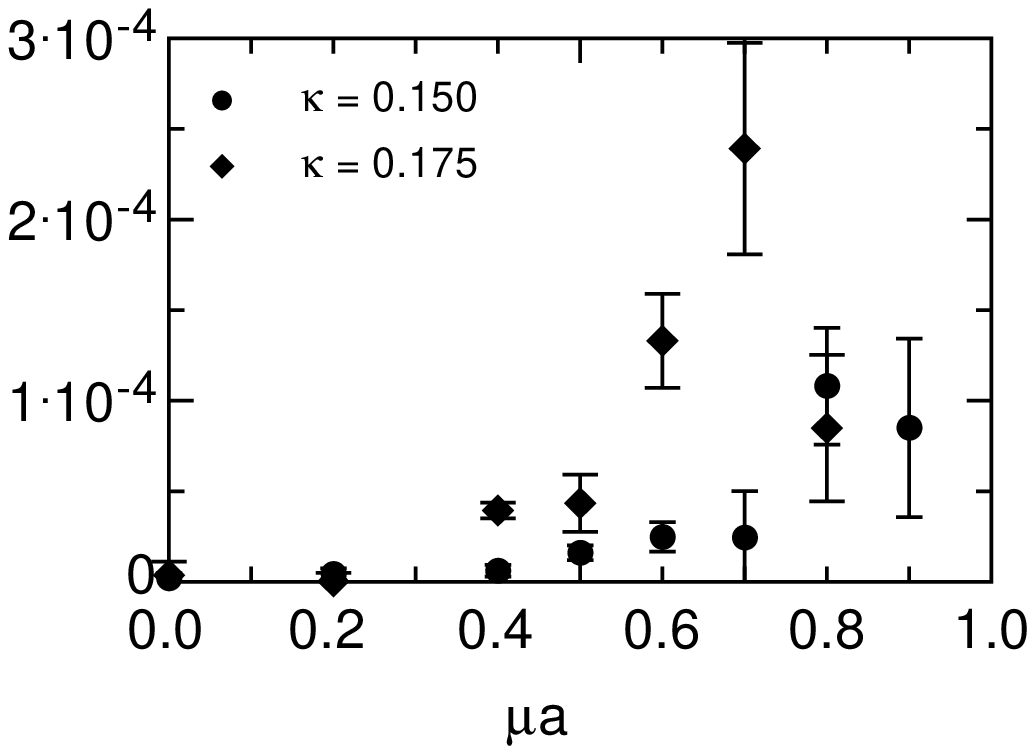}
\end{minipage}
\begin{minipage}{ 0.4\linewidth}
\includegraphics[width= \linewidth]{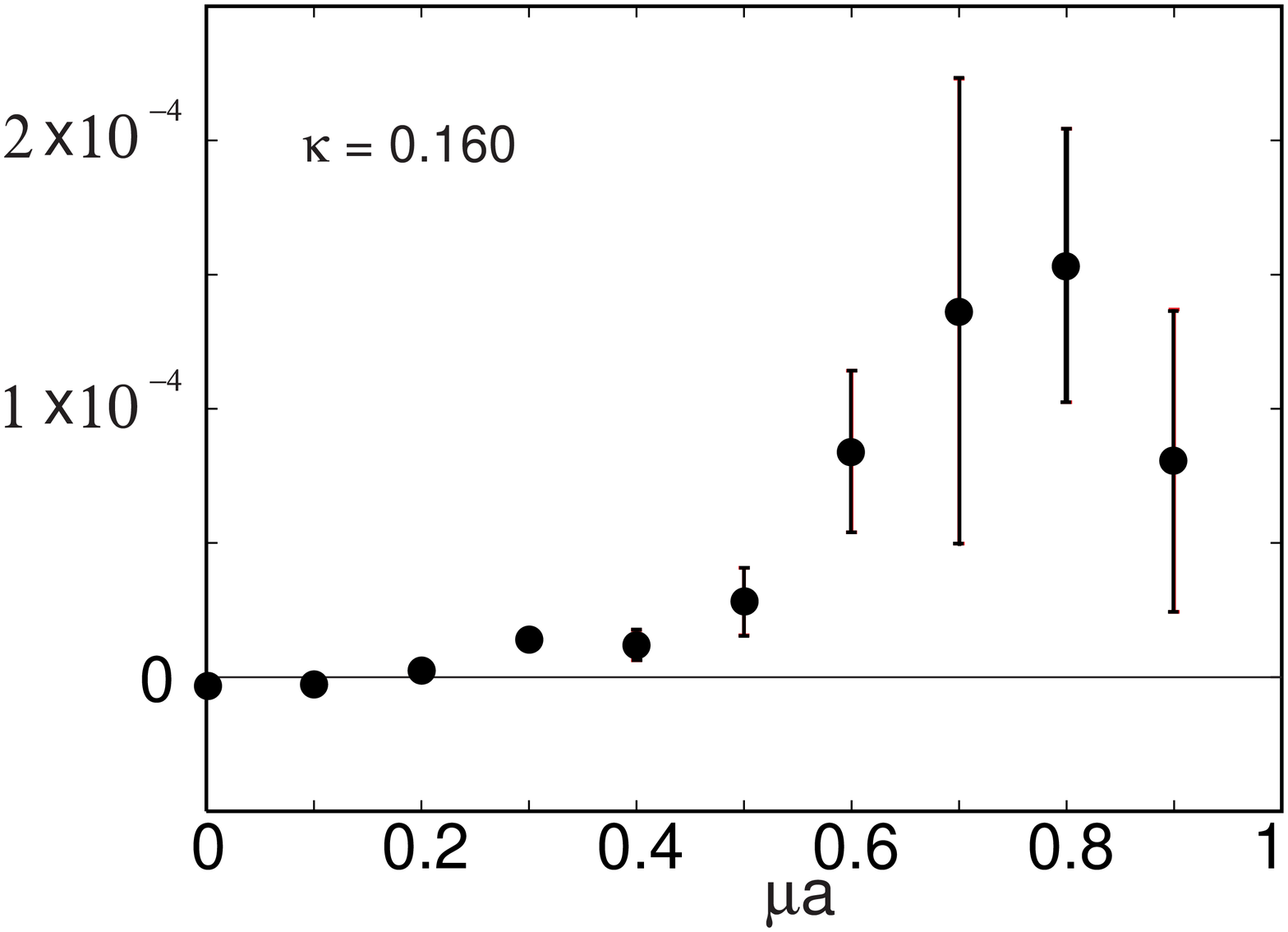}
\end{minipage}
\end{center}
\vspace{-8mm}
\caption{Polyakov line susceptibility,
$\partial \langle L \rangle/\partial/\mu = 
\langle (L-\langle L \rangle)(n-\langle n \rangle) \rangle$,
as a function of $\mu$ on $4^3\times 8$ lattice. 
(a) $\kappa=0.150,0.175$ under the anti-periodic spatial boundary
condition 
and
(b) $\kappa=0.160$ under the periodic boundary
condition. 
}
\label{Fig-PolSus}
\end{figure}

In Fig.\ref{Fig-PolSus}, we show the Polyakov line susceptibilities 
$\kappa=0.150$ and $0.175$ under the antiperiodic spatial boundary
condition and for $\kappa=0.160$ under the periodic boundary
condition. 
All exhibit a peak when $\mu$ increases, which indicates
a deconfinement transition.
All quantities support the picture that at large $\mu$  the system
undergoes the transition from the confinement to the deconfinement phase. 
In addition to
the increase of $\langle L \rangle$ and $\langle E_g\rangle$, the rapid increase of $\langle n \rangle$ is 
observed. The instability at large $\mu$ may be an indication of
a new phase with the diquark condensation.
\footnote{
Since our fermion action includes only bilinear terms of $\bar{\psi}$
and $\psi$ and not those of $\psi$ and $\psi$, or 
$\bar{\psi}$ and $\bar{\psi}$, the diquark condensation cannot emerge. 
}

Although the behavior of all quantities supports the existence of the 
deconfinement phase at large $\mu$, there are some indications
that suggest a more complicated phase. In many cases, we observe a
second peak in the Polyakov line susceptibility at large $\mu$.

\section{Hadron propagators}

We calculate correlations of color singlet hadron operators,
$M(x)=\bar{\psi}^a_\alpha(x)\Gamma_{\alpha\beta}\psi^a_\beta(x)$ 
and $B(x) = \epsilon^{ab}\psi^a_\alpha(x)(C\Gamma)_{\alpha\beta}
\hat{\tau}\psi^b_\beta(x)$,
where  $\Gamma$ is the product of Dirac matrices and 
$\hat{\tau}$ is a Pauli matrix acting on flavor indices.
$C$ is the charge conjugation matrix and $a$ and $b$ are color
indices. 
We set $\hat{\tau}=\tau_2$ or $\tau_2 \vec{\tau}$ 
so that the wave function is totally antisymmetric.

To our knowledge, 
no study of the behaviors of hadrons including vector mesons 
at finite baryon density has been performed using lattice QCD.
Vector mesons are important since they provide information at
several stages of heavy ion collision in the form of lepton pairs.

\begin{figure}[hbt]
\begin{center}
\begin{minipage}{ 0.48\linewidth}
\includegraphics[width=1.0 \linewidth]{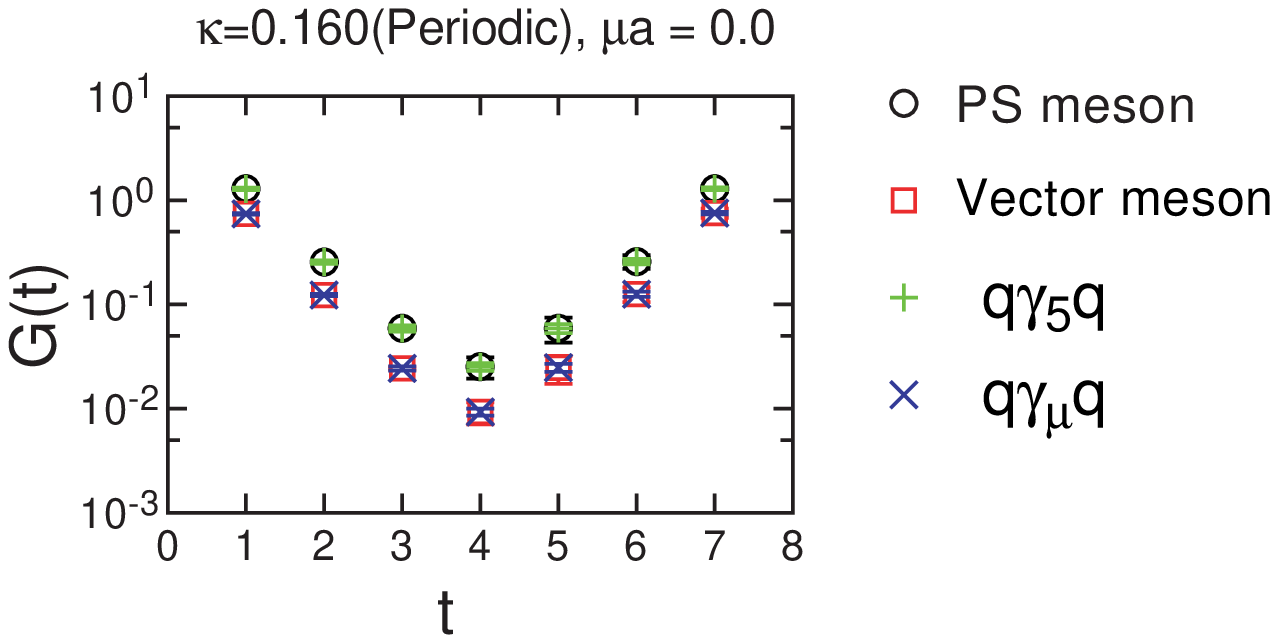}
\end{minipage}
\hspace{1mm}
\begin{minipage}{ 0.48\linewidth}
\includegraphics[width=1.1\linewidth]{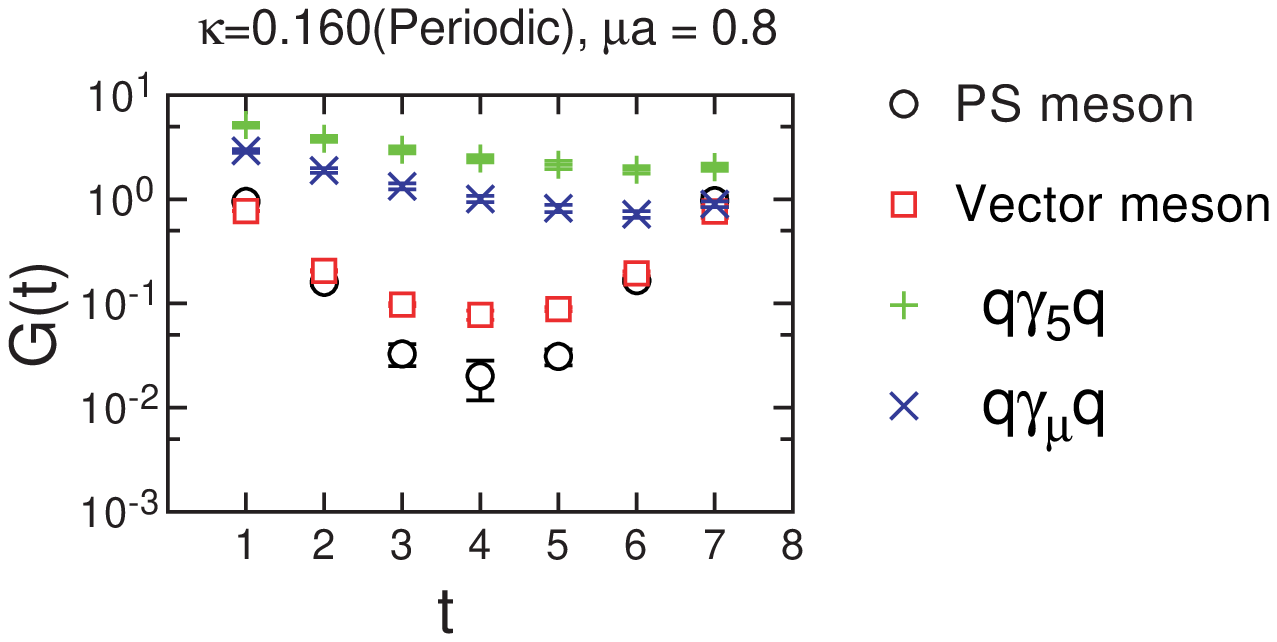}
\end{minipage}

\end{center}
\caption{
Hadron propagators at $\mu a=0$ (left) and $\mu a=0.8$.
Pseudoscalar and vector mesons and scalar and pseudovector
baryons are shown.
}
\label{Fig-propa}
\end{figure}

\begin{figure}[hbt]
\begin{center}
\begin{minipage}{ 0.48\linewidth}
\includegraphics[width=1.0 \linewidth]{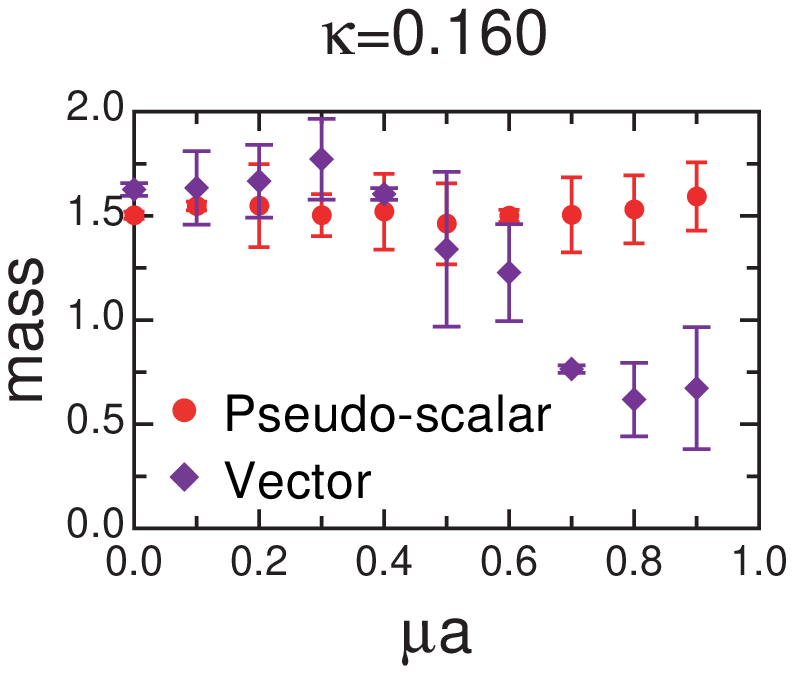}
\end{minipage}
\hspace{1mm}
\begin{minipage}{ 0.48\linewidth}
\includegraphics[width=1.1\linewidth]{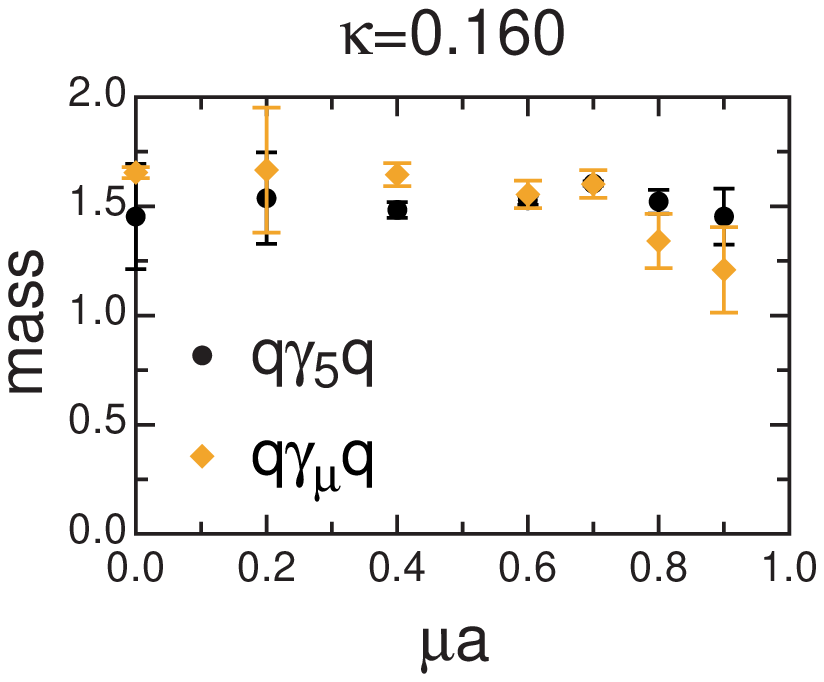}
\end{minipage}

\end{center}
\caption{
Meson mass (left) and diquark mass (right) as a function of $\mu$
at $\kappa=0.160$ with the periodic boundary condition.
}
\label{Fig-pirho}
\end{figure}

In Fig.\ref{Fig-propa}, we show propagators of the pseudoscalar and
vector mesons and those of the scalar ($\Gamma=\gamma_5$) and pseudovector
($\gamma_\nu$) baryons for $\mu=0$ and $\mu a =0.8$, where $a$ is the lattice
spacing.  
At $\mu=0$, the propagator of pseudoscalar mesons is equivalent to that of
scalar baryons, 
and that of the vector meson is equivalent to that of the pseudovector baryon. 
We see that this relation is satisfied in the numerical calculation.
The most prominent feature here is that the vector meson propagator is
strongly modified at $\mu a=0.8$.  
Its slope is more gradual than that for the pseudoscalar,
i.e., the vector meson becomes lighter.

The lattice size of $4^3\times 8$ is too small to extract information
of the mass pole. 
Nevertheless, we fit the data to clarify the chemical
potential effect qualitatively, and 
in Fig.\ref{Fig-pirho} 
we plot the scalar and vector meson masses together with those of corresponding
baryons,
as a function of the chemical potential.
We find that the vector meson mass drops as $\mu$ reaches the
critical region.  In order to confirm this unexpected result, we
calculate both periodic and antiperiodic boundary conditions and
several $\kappa$'s ($\kappa=0.150, 0.160, 0.175$);
the reduction of vector meson mass is always observed.
Although our lattice size is too small and the statistics
are insufficient for extrapolation to the chiral
limit, the signal of the anomalous behavior of the vector channel 
is clearly seen.

At $\mu=0$, because of the QCD inequality \cite{KSteT}, 
the lightest mass should be in the pseudoscalar channel.
The inequality holds under two assumptions, i.e.,  (i)
there is no disconnected diagram and (ii) Eq.(\ref{D-real}) is valid. 
The second condition is not satisfied for the finite density state.
Indeed, there are several conjectures in the literature.
Brown and Rho first proposed the scaling law 
$m^*/m=f_\pi^*/f_\pi$
to explain the large low mass lepton pair enhancement 
observed in CERES \cite{CERES}, 
where $m^*$ and $f_\pi^*$
are the mass and the pion decay constant in the medium \cite{BR-theo}.
Based on the QCD sum rule, Hatsuda and Lee predicted a decrease of
$\rho$ meson mass as a function of $\mu$ \cite{H-L}.
Harada et al. showed that vector meson mass vanishes at the
critical density as a consequence of an effective theory 
with hidden local symmetry \cite{Harada}.
Yokokawa et al. proposed simultaneous softening of $\sigma$ and 
$\rho$ mesons associated with the chiral restoration
\cite{Yokokawa}.
If the sudden drop of the vector meson mass is not a special
feature of the color SU(2) model, this may be the 
first lattice QCD result to show the reduction of the vector meson
mass in the medium.

We do not observe any special feature in baryon (diquark) channels.
We fit the baryon propagators as
\begin{equation}
G(t) = C_1 e^{-(m-2\mu)t} + C_2 e^{-(m+2\mu)(N_t-t)},
\end{equation}
where $C_1$ and $C_2$ are not independent because of the
boundary condition, $G(0)=G(N_t)$.
At $\mu > 0$, the pseudovector baryon propagator is not equivalent to
the vector meson, but their masses are similar.

\section{Concluding remarks}

In this study, we have investigated the Polyakov line and its 
susceptibility together with thermodynamical quantities
for Wilson fermions in $(\kappa,\mu)$ parameter space. 
Since there has been no lattice study using Wilson fermions
and improved gauge action for the two-color finite density system, 
this step is necessary in order to understand in which region 
hadron propagators in the medium change their behavior.

We  observed a sudden reduction of the vector meson mass when
the chemical potential was increased.  If this is due to the mixing
of meson and diquark states, the mass of diquark partner, 
$\psi_c\gamma_\mu \psi$,
should increase, but this is not the case.  
Rather, $\psi_c\gamma_\mu \psi$ behaves
in a similar way to the vector.
As an urgent project, we will perform simulations on larger lattices,
which will allow us to perform the chiral limit extrapolation 
to estimate the physical scale.

Our lattice here is small, 
but results for the case of periodic  and antiperiodic boundary 
conditions in the spatial directions show the same qualitative behavior.

The behavior of the thermodynamic quantities together with that of the
Polyakov line supports the standard picture, i.e., QCD undergoes a transition
from the confinement to the deconfinement phase.
We have observed several indications in the susceptibility and
gluon propagators, which may suggest a more complicated
phase at finite baryon number density.
The number density may play an essential role in confirming the situation;
it should increase from $\mu=M_B/N_c$, where
$M_B$ is the lightest baryon mass and $N_c$ is the number of colors.
At the deconfinement transition point, $\mu_c$, particle which bears
the baryonic charge is changed from hadrons to quarks, therefore
the mass of particles corresponding to the chemical potential
changes. This results in the jump of the number density
at $\mu_c$.
In the following paper, 
we will perform detailed study of $\langle n \rangle$ and its susceptibility.

\section*{ Acknowledgment}

We would like to thank T. Kunihiro, T. Inagaki and T. Sch\"afer for helpful
discussions.
This work is supported by Grant-in-Aide for Scientific Research by
Monbu-Kagaku-sho (No.11440080, No. 12554008 and No. 13135216),
and ERI, Tokuyama Univ..
Simulations were performed on SR8000 at IMC, Hiroshima
M
Univ., SX5 at RCNP, Osaka Univ., SR8000 at KEK.


\end{document}